\documentclass[aps,preprint,showkeys,nofootinbib,prb]{revtex4-1}%
\usepackage{amssymb}
\usepackage{amsfonts}
\usepackage{amsmath}
\usepackage{graphicx}
\usepackage{natbib}%
\providecommand{\U}[1]{\protect\rule{.1in}{.1in}}
\setcitestyle{numbers}
\setcitestyle{square}

\begin{document}
\title{On pseudogap phase as precursor to a superconducting dome in high-T$_{C}$
cuprates: Non-analytic $T^{\ast}$ as a function of doping }
\author{Felix A. Buot$^{\ast}$}
\affiliation{C\&LB Research Institute, Mount Tugas, Carmen 6005, Cebu, Philippines.}
\affiliation{$^{\ast}$Email: felixa.buot@gmail.com}

\begin{abstract}
We generalize the condition under which a quantum material exhibiting a
pseudogap phase is a precursor to a superconducting (SC) dome. The result
reveals the non-analytic $T^{\ast}$ as a function of doping. A well-known
example is the high-$T_{C}$ cuprates. Essentially, the SC dome is generated
under two conditions: (1) that the pseudogap $T^{\ast}$ is a decreasing
function of doping, due to the decrease in size of extended pairing of doped
holes with doping, and most importantly, (2) that the rate of
configurational-ordering parameter is an increasing function of doping as a
result of the decrease in extended length of the disordered pairs. The theory
behind these hinges on a novel strong entanglement and confinement hole
pairing (ECHP) mechanism that unravels the microscopic features of the entire
phase diagram of both electron and hole-doped high-$T_{C}$ cuprates, recently
discussed in the by Buot et al. The ECHP configurational ordering (CO) at the
underdoped and overdoped regions qualitatively agrees with spin texture
experiments, and strongly suggests that strange metal phase can only occur in
the overdoped region, in agreement with experiments.

\end{abstract}
\maketitle

\section{Introduction}

The purpose of this brief communication is to demonstrate that any
superconducting materials exhibiting a pseudogap phase, owing to disordered
\textit{extended} preformed "Cooper" pairs of variable confinement or coupling
strength as a function of doping, must be a precursor to the formation of the
SC dome phase. The variable confinement implies \textit{variable length of}
\textit{extended entangled pairs} as a function of doping. Because of
difference in pair extended sizes (difference in reordering transition
energy), there is an inherent dependency of the rate of approach to a
symmetry-breaking phase leading to superconductivity. These conditions are
satisfied by the new pairing mechanism, introduced by Buot, Otadoy, and Pili
\cite{bop} which we will refer to as BOP in what follows. The new pairing
mechanism is based on entanglement and confinement pairing of doped holes.
This new pairing mechanism and its role in explaining the microscopic theory
of high-T$_{C}$ cuprates is discussed in more details by BOP.

So far, most discussions in the literature focuses only on the pseudogap phase
often yielding an entire analytical or continuous behavior for the $T^{\ast}$
as function of doping. These discussions disregard the SC dome as separate
from the calculation of $T^{\ast}$. Here, we found that the SC dome is tightly
coupled to the behavior of $T^{\ast}$ and turns out to be a derivative of the
pseudogap phase. The entire analyticity of $T^{\ast}$ no longer holds. Indeed,
the behavior of $T^{\ast}$ is non-analytic. This emerges when the
configurational-order rate goes infinite at the peak of the SC dome and all
through the overdoped region. As we shall see, this also suggests that strange
metal phase occurs only in the overdoped region.

\section{Macroscopic generation of the SC dome}

\begin{figure}
[ht!]\centering\includegraphics[width=5.785900in]{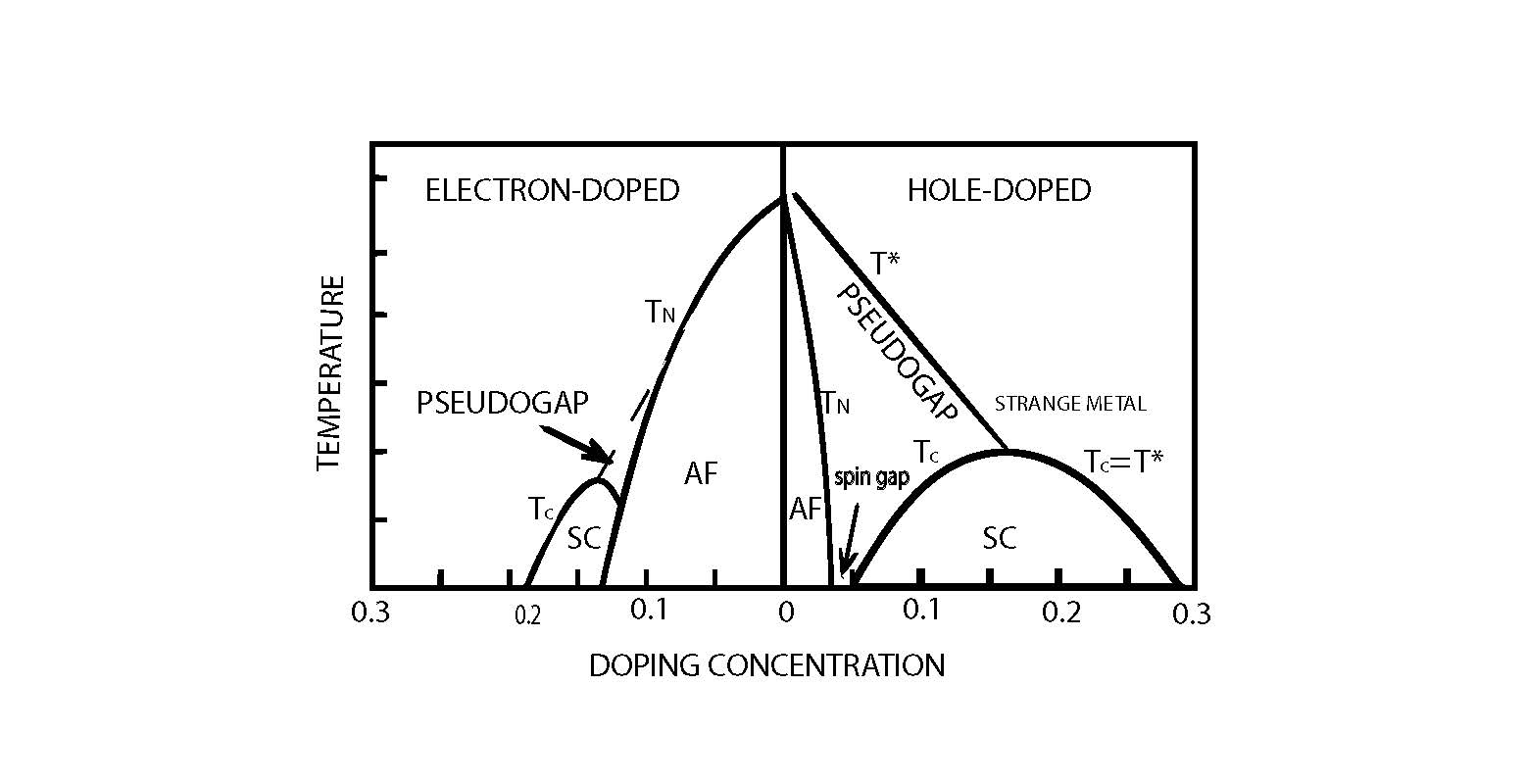}
\caption{Generic phase diagram for the high $T_{C}$ superconductors. The underdoped region is defined to be the doping range below the peak of the SC dome, whereas the overdoped region are doping levels beyond the peak. [Figure redrawn and edited from Ref. \cite{edeger}]}
\label{phaseD}
\end{figure}

The macroscopic physics of this phase diagram is mainly characterized by two
order parameters, namely, (1) the \textit{local} phase ordering of
configurational-disordered preformed \textit{pair order} (PO), that is,
condensation to lower energies at $T^{\ast}$ of the disordered "nematic"
extended preformed entangled pairs and (2) the \textit{global} "smectic"
\textit{configurational order} (CO) that is a symmetry-breaking (SB)
transition from "nematic" to "smectic" ordering at $T_{C}$ of the SC dome,
eventually adhering to $T^{\ast}=T_{C}$ in the overdoped region. The spin gap
phenomenon is an exception to this rule since it fails to attain a CO, where
the configurational arrangement becomes unique. An auxiliary order parameter
describes the rate of evolution from the \textit{local} phase to
\textit{global} symmetry-breaking phase. The typical phase diagram of cuprates
is shown in Fig. \ref{phaseD} taken from Edegger et al \cite{edeger}. This is
supported by ARPES experiments \cite{damascelli}.

Since the spatial arrangement of extended preformed pairs are distributed
randomly in the vertical and horizontal directions, we may characterize this
situation with one variable, namely, the configuration entropy ($CE$). This
seems a convenient single quantity to characterize the evolution of disordered
spatial configurations of extended preformed pairs as the temperature is
further lowered from PO at $T^{\ast}$. Clearly, a null $CE$ signifies that the
spatially-disordered configurations of extended preformed pairs undergo
symmetry-breaking \textquotedblleft smectic\textquotedblright\ configurational
transition to a definite unique rearrangement, and at CO we have,
\begin{equation}
CE=Log\ \left(  1\right)  =0 \label{logzero}%
\end{equation}
at the SC dome. \ At $CE=0$, the CO phase consists of parallel superconducting
stripes, as derived qualitatively by BOP.

The temperature dependence of configurational entropy $(CE)$ or the evolution
of $CE$ as the temperature is lowered from $T^{\ast}$ is simply assume to have
a linear temperature dependence. The slope, $R$, of this linear function,
lines ($CE\left(  T\right)  ^{\prime}s$) emanating from each $T^{\ast}$-point,
indicate the \textit{rate of evolution} towards the superconductivity phase
($CE=0$) as the temperature is lowered. We emphasize that this slope, $R$, is
proportional to the doping level as well, in recognition of the variable
length of disordered extended pairings. Clearly, shorter extension of
disordered pairs, resulting from increase in doping levels, is more
"fluid"(lower phase transition energy) and is therefore expected to have
larger $R$ than for longer extended disordered pairs. Thus, our
configuration-ordering rate parameter is given by%
\begin{equation}
R\left(  d\right)  =\left.  \frac{\partial CE\left(  T\right)  }{\partial
T}\right\vert _{doping=d} \label{rigid}%
\end{equation}
The rate $R\left(  d\right)  $ is a measure of the \textquotedblleft
flexibility\textquotedblright\ of the random arrangements of preformed pairs
to evolve towards symmetry-breaking \textquotedblleft
smectic\textquotedblright\ CO phase transition, where Eq. (\ref{logzero})
holds at the SC dome. We should point out that in RVB theory, the Anderson
ordering parameter was also taken to have a linearly increasing dependence
with temperature and doping levels \cite{singh2}.

The evolution process from $T^{\ast}$(PO) to $T_{C}$ (CO) as the temperature
is lowered is thus described by the transition rate $R$. Clearly, at CO,
$CE=0$, since CO signifies a unique arrangement of entangled pairs. The spin
gap is characterized by the failure to attain CO upon continued cooling from
$T^{\ast}$ down to a limiting lowest temperature because the rate $R$ is
insufficient to attain CO. On the other hand, at the peak of the SC dome $R$
goes to infinity, indicating a non-analyticity of $T^{\ast}$. It is worth
emphasizing the singularity of $T^{\ast}$at the SC dome peak, and the
coincidence of $T^{\ast}=T_{C}$ in the overdoped region of the SC dome are
consistent with experimental result typified by Fig. \ref{phaseD}.

\begin{figure}
[ht!]\centering\includegraphics[width=6.785900in]{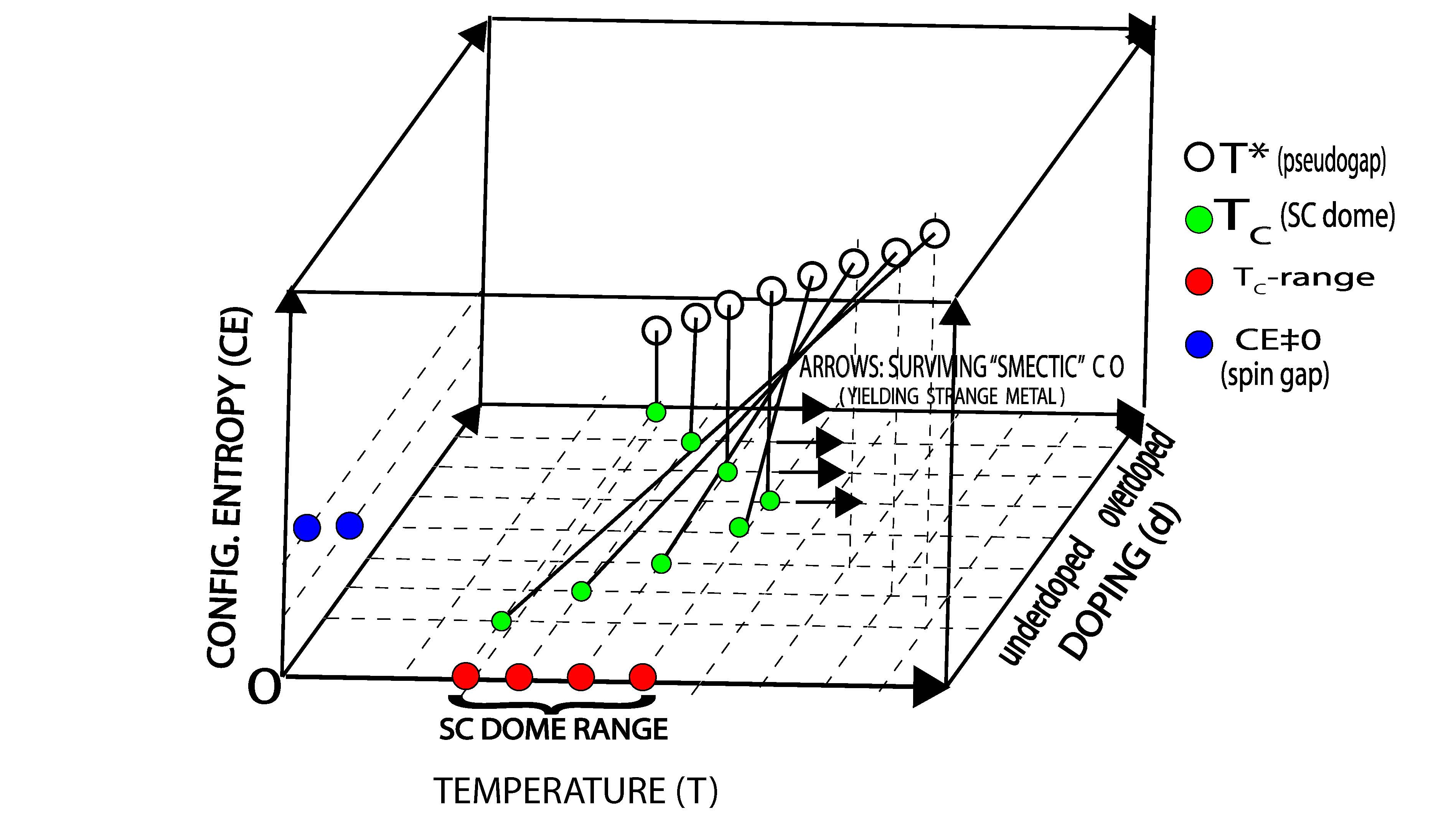}
\caption{The plot is compatible with decrease in $T^{\ast}$ with increase in doping and decrease in temperature, in agreement with the experimental behavior in Fig.\ref{phaseD}. The temperature drop from $T^{\ast}$ to $T_{C}$ is inversely proportional to the slope or rate R of the lines joining these points. The lines joining the spin gap points are not shown. The arrow denotes the surviving CO at temperature above SC dome, maintaining the one-dimensional conducting stripes obeying the laws
of Planckian mesoscopic physics \cite{bop}. This yields a linear T-dependence
of resistivity of the strange metal.}\label{schematic1}
\end{figure}

In Fig. \ref{schematic1}, the $CE$ vs. $T$ relation is assume to obey linear
evolution from $T^{\ast}$ with the rate $R$ increasing with increase of doping
levels. This corresponds to the linearly increasing Anderson order parameter
with doping levels in the RVB theory. The singularity occurs where the rate
$R$ goes to infinity, manifested initially as a "kink" of $T^{\ast}$ at the
peak of the SC dome. $R$ goes to infinity all the way down through the entire
overdoped region of the SC dome. At the overdoped region of the SC dome phase,
$T^{\ast}=T_{C}$. Clearly, the behavior of $T^{\ast}$ is entirely
non-analytic. The arrows denote the surviving $CE=0$ at $T>T_{C}$ above the SC
dome at the overdoped region, maintaining the parallel one-dimensional
normally-conducting stripes obeying the laws of Planckian mesoscopic physics
\cite{bop}. This yields a linear $T$-dependence of resistivity of the strange
metal in accordance with mesoscopic physics in 1D.

\begin{figure}
[ht!]\centering\includegraphics[width=6.785900in]{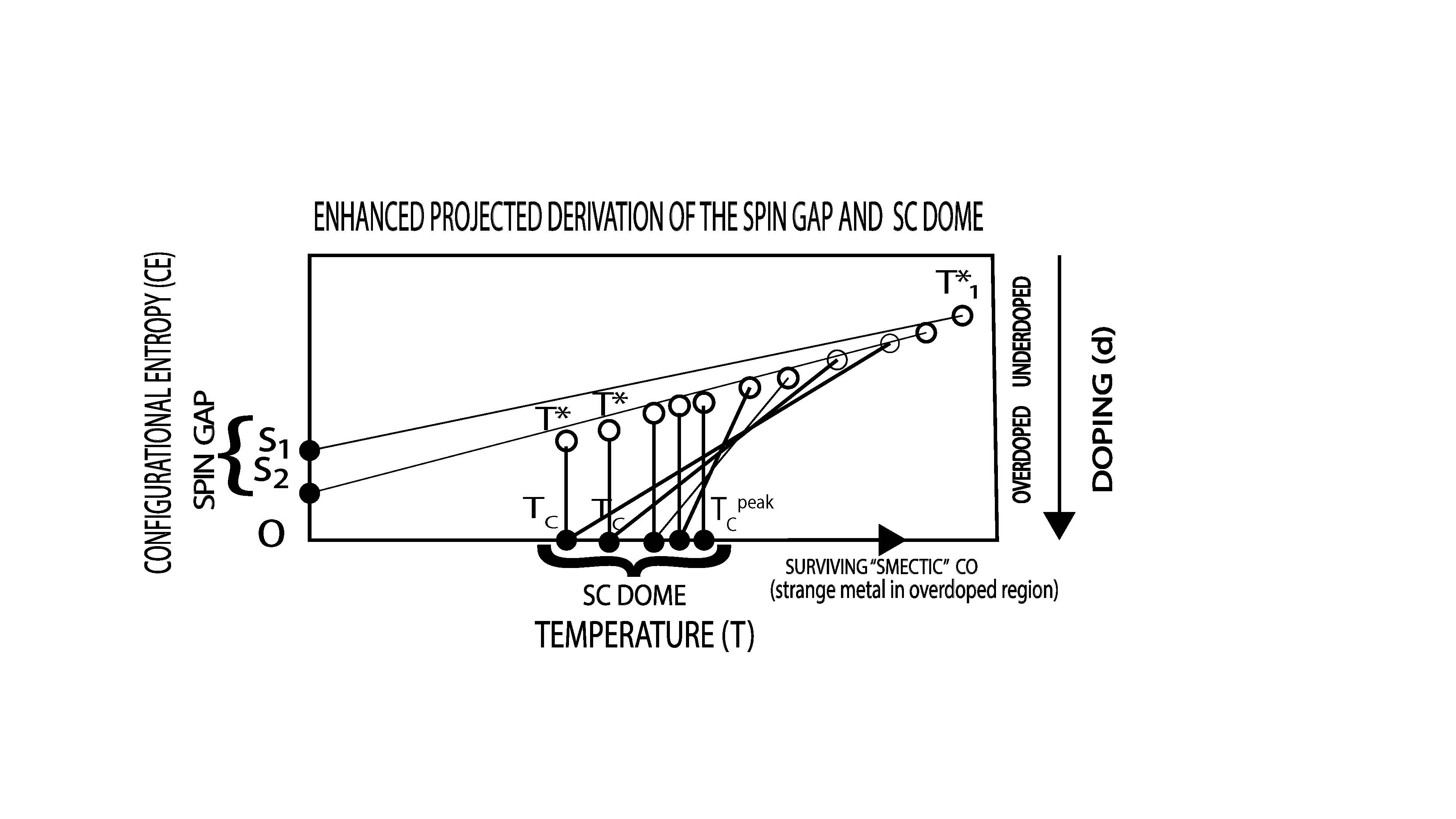}
\caption{This figure is an enhanced projected view from Fig.\ref{schematic1}
intended to emphasize the spin gap, SC dome and conditions leading to strange metal behavior. The singularity occurs where the rate R goes to infinity. It retraces the locus (black dots) of the CO's produced by $T^{\ast
}$ in the underdoped region thus forming the SC dome phase where $T^{\ast
}=T_{C}$.The arrow denotes the surviving CO at temperature above
SC dome, maintaining the one-dimensional conducting stripes obeying the laws
of Planckian mesoscopic physics \cite{bop}. This yields a linear T-dependence
of resistivity of the strange metal.}\label{schematic2}
\end{figure}

Thus, without sophisticated many-body calculations, with a knowledge of the
experimental data on the pseudogap phase as a decreasing function of doping,
one only need some insights into the configurational-ordering rate of
disordered extended pairings as an increasing function of the doping levels to
deduce the so-called spin gap and SC dome, as discussed above. This is readily
understood through the graphical generation of SC dome of Fig.
\ref{schematic1}. Observe that the drop in temperature from the pseudogap
$T^{\ast}$ to SC $T_{C}$ is inversely proportional to the slope of the lines
connecting them, denoted by the rate $R$. This brings about a retracing on the
temperature axis of the $T_{C}$ produced by lines with finite slopes by the
$T_{C}$ of the lines with infinite slopes giving rise to the SC dome, in
agreement with experimental data of Fig. \ref{phaseD}. For infinite slopes,
$T^{\ast}$ and $T_{C}$ are identical as shown in Fig. \ref{schematic1} and
verified by the experimental phase diagram of Fig. \ref{phaseD}. In Fig.
\ref{schematic1}, the lines joining the spin gap points are not shown, these
lines are exhibited in a simulated projected view of Fig. \ref{schematic1},
this is shown in Fig. \ref{schematic2}.

The arrow along the temperature axis in Fig. \ref{schematic2}, also portrayed
in Fig. \ref{schematic1} denotes the surviving $CE=0$ at $T>T_{C}$ at
temperature above SC dome in the overdoped region, maintaining the
one-dimensional normally-conducting stripes obeying the laws of Planckian
mesoscopic physics \cite{bop}. This yields a linear T-dependence of
resistivity of the strange metal. Figure \ref{schematic2} is intended to
emphasize the spin gap and its duality with the strange metal behavior,
namely, for spin gap $CE\neq0$, $T<T_{C}$, whereas, for strange metal, there
is a surviving $CE=0,$ at $T>T_{C}$ which leads to strange-metal behavior
\cite{bop}.

\section{The need for an unconventional pairing theory}

We emphasize that we were able to generally demonstrate that a pseudogap phase
is inherently a precursor to the formation of a SC dome. In macroscopic
generation above, we do not need to know the microscopic pairing mechanism,
only that these are disordered \textit{extended} preformed pairs in the
pseudogap phase. We use only discernible data, namely, the knowledge of the
experimental $T^{\ast}$, as well as insights into the configurational-ordering
rate $R$.

However, to go deeper into the physics, a new unconventional strong pairing
theory is needed to justify the assumptions of the graphical generation of the
SC dome above, among many other unexplained features of cuprates. This is a
theory where the coupling, $\Delta$, and hence $T^{\ast}$, decreases with
doping levels and that configurational-ordering rate $R$ increases with doping
levels. The needed theory must explain the experimentally-found parallel
conduction stripes on the CuO plane of cuprates as well as the linear
resistivity of the so-called strange-metal phase. Moreover, the needed theory
must also explain the experimental spin textures of some cuprate quantum SC
materials. These are the microscopic features that need to be understood in
the light of the these experimental observations. These issues are
qualitatively discussed in greater details by BOP and interested readers are
referred to this reference.

\section{Concluding Remarks}

The SB "smectic" CO current-carrying pattern depicted in Figs.7 and 8 of BOP
can readily explain the rivers of superconductive charge, spin-polarized and
spin-unpolarized stripes, as natural consequences of entanglement and
confinement hole pair (ECHP). The idea of confinement and extended pairing
also helps to elucidate the decrease in $T_{C}$ with overdoping, as well as
the decrease in $T^{\ast}$ in the pseudogap phase, the discernible increase in
$R$ with doping levels, as well as the $T^{\ast}$-"kink" at the SC dome peak,
Figs. \ref{schematic1} . The SB "smectic" CO which persists at $T>T_{C}$ in
the overdoped region predicts a linear-$T$ resistivity of 1D mesoscopic
physics of stripes. For $T>T_{C}$, the holes at both ends of the
antiferromagnetic link no longer move in unison with charge 2%
%TCIMACRO{\TEXTsymbol{\vert}}%
%BeginExpansion
$\vert$%
%EndExpansion
e%
%TCIMACRO{\TEXTsymbol{\vert} }%
%BeginExpansion
$\vert$
%EndExpansion
but now move in an uncoordinated manner as independent 1D parallel channels
\cite{bop}. The spin texture experiments have also been qualitatively
discussed in terms of the new theory by BOP.

In summary, the entanglement and confinement pairing model qualitatively
explains the entire phase diagram of cuprates, both the macroscopic and
microscopic physics in the light of experiments \cite{bop}. That the pseudogap
phase as a precursor to the SC dome is clarified in more detail in this
communication, as well as the duality of spin gap phase ($CE\neq0$) and
strange metal ($CE=0$ only) phase, which can occur only in the overdoped
region of the phase diagram. The superconducting stripes, and strange-metal
linear-$T$ behavior above the overdoped regions of SC dome \cite{over} is
treated in greater detail by BOP. The strange metal behavior at $T>T_{C}$ in
the overdoped region maybe also be attributed to the influence of the in-plane
electric fields in maintaining "smectic" $CE=0$ at $T>T_{C}$ with doping
concentration. The readers is referred to BOP for greater details.

\begin{figure}
[ht!]\centering\includegraphics[width=6.785900in]{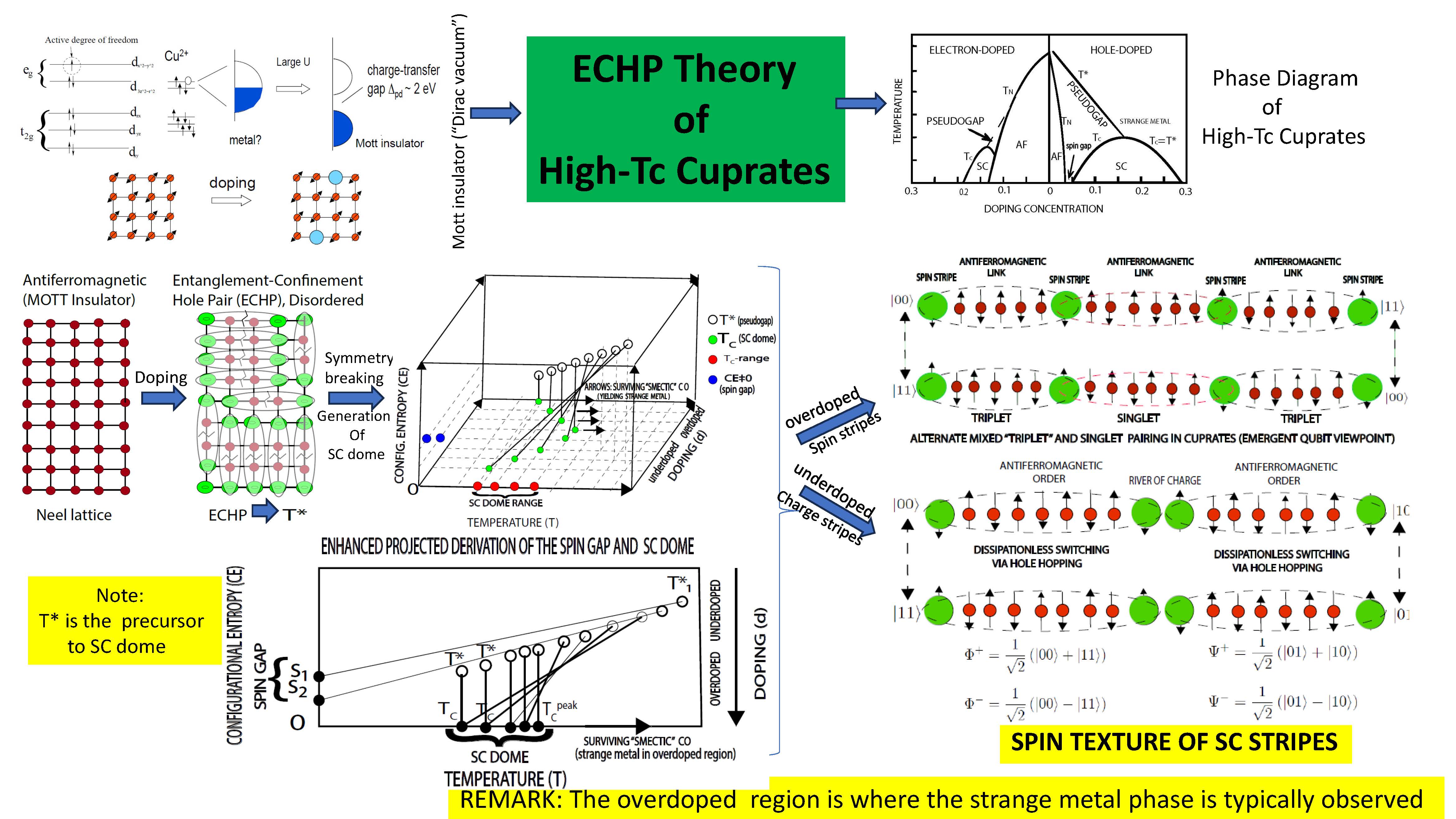}
\caption{A graphical outline of the ECHP theory of high-T$_{C}$ cuprates}\label{figsummary}%

\end{figure}

A summary of the ECHP theory is shown graphically in Fig. \ref{figsummary}. As
shown in the graph, the CO at overdoped region looks like a "global
interlocking" of all ECHP's \ giving a CO "rigidity" that survives at
$T>T_{C}$, yielding the strange metal phase.

\begin{thebibliography}{9}                                                                                                %


\bibitem {bop}F.A. Buot, R.E. S. Otadoy, and U. Pili, \textit{Entanglement and
confinement: A new pairing mechanism in high-T}$_{C}$\textit{ cuprates},
arXiv:2410.02300v5 [cond-mat.supr-con] 02 Nov 2025. Referred to as BOP in the text.

\bibitem {edeger}B. Edegger, V. N. Muthukumar and C. Gros,
\textit{Gutzwiller--RVB theory of high-temperature superconductivity: Results
from renormalized mean-field theory and variational Monte Carlo calculations},
Adv. Phys. \textbf{56}, (6), 927--1033 (2007).

\bibitem {damascelli}A. Damascelli, Z.-X. Shen, and Z. Hussain,
\textit{Angle-resolved photoemission spectroscopy of the cuprate
superconductors}, Rev. Mod. Phys. \textbf{75}, 473 (2003).

\bibitem {singh2}N. Singh, \textit{Leading theories of the cuprate
superconductivity: a critique}, arXiv:2006.06335v2 (2020).

\bibitem {over}N.E. Hussey, H. Gordon-Moys, J. Kokalj and R.H. McKenzie,
\textit{Generic strange-metal behaviour of overdoped cuprates}, J. Physics:
Conference Series \textbf{449}, 012004 (2013).
\end{thebibliography}
\end{document}